# Quantum Cellular Neural Networks

*Geza Toth, Craig S. Lent, P. Douglas Tougaw,*

*Yuriy Brazhnik, Weiwen Weng, Wolfgang Porod*

*Ruey-Wen Liu, Yih-Fang Huang*

*Department of Electrical Engineering*

*University of Notre Dame*

*Notre Dame, Indiana*

**Abstract**

We have previously proposed a way of using coupled quantum dots to construct digital computing elements - quantum-dot cellular automata (QCA). Here we consider a different approach to using coupled quantum-dot cells in an architecture which, rather that reproducing Boolean logic, uses a physical near-neighbor connectivity to construct an analog Cellular Neural Network (CNN).

## I. Introduction

We discuss a computing paradigm in which cells composed of interacting quantum dots are employed in a cellular neural network (CNN) architecture. Communication between cells is only through the Coulomb interaction. The cells and their basic behavior are the same as we have previously discussed in the context of the Quantum-dot Cellular Automata (QCA) architecture. The key differences here are that in the quantum CNN (Q-CNN) approach: (1) Each cell is used to encode a continuous rather than binary degree of freedom. (2) We focus on the time dependent problem instead of the ground state. (3) The time-dependent Schrödinger equation can be transformed into the CNN state equations.

We have constructed a simple quantum model of a Q-CNN composed of quantum-dot cells. Each cell contains one classical degree of freedom, the cell polarization, and one quantum degree of freedom, a quantum mechanical phase difference. Mapping onto the CNN paradigm maintains phase information within the cell but no quantum coherence exists between cells. Thus though dynamics is accomplished through the quantum degrees of freedom, information is only carried across the array in classical degrees of freedom.

Our hope is that by connecting the problem of coupled quantum cells to a circuit architecture developed for exploiting conventional analog integrated circuits, we might be able to open up a new solution domain for interconnected quantum devices. Because local connectivity is natural in ultra-small quantum devices, CNN's may prove a natural extension to the QCA architecture and allow a move into non-digital domains.

In Section II we briefly review the CNN paradigm. In Section III a quantum treatment of a cellular array will be described. In Section IV the connection between the quantum problem and the CNN approach will be demonstrated. In Section V. we discuss the generalization of our simple model to a more general class of Q-CNN's.

## II. The CNN paradigm

The CNN, invented by L. O. Chua and L. Yang [2] and generalized in subsequent work [3,4], is a two or three dimensional, usually regular array of analogous cells. Each cell, indexed by $\kappa$, has dynamical state variables $\vec{x}_\kappa$, external inputs $\vec{u}_\kappa$, and internal constant cell data $\vec{z}_\kappa$. Each cell is influenced by its neighbors through a synaptic input $I^s_\kappa$ which depends on the values of cell states and cell inputs within a sphere $S_\kappa$ centered on cell $\kappa$. A *CNN synaptic law* describes the effect of other cells on the synaptic input.

$$I^s_\kappa = \sum_\lambda A^\lambda_\kappa \vec{x}_{\kappa+\lambda} + \sum_\lambda B^\lambda_\kappa f(\vec{x}_\kappa, \vec{x}_{\kappa+\lambda}) + \sum_\lambda C^\lambda_\kappa \vec{u}_{\kappa+\lambda} \qquad (1)$$

The cell dynamics are determined by a *CNN state equation* giving the rate of change of state variables as the *nonlinear* function of the state of the cell itself, the synaptic input from neighboring cells, and the external inputs.

$$\frac{\partial}{\partial t} \vec{x}_\kappa = -g(\vec{x}_\kappa, \vec{z}_\kappa, \vec{u}_\kappa, I^s_\kappa) \qquad (2)$$

If the no external inputs exist then the CNN is called *autonomous*. The CNN is then defined by (1) the synaptic law, (2) the state equation, (3) initial conditions, and (4) boundary conditions. Unlike neural networks in case of the CNN the cells are primarily *locally interconnected*, thus the practical realization is much easier, than in the case of a fully interconnected neural network.

## III. Quantum model of cell array

We consider here a simple model of an array of interacting quantum cells. Each cell contains four quantum dots and two extra electrons as shown schematically in Figure 1. The electrons tend to

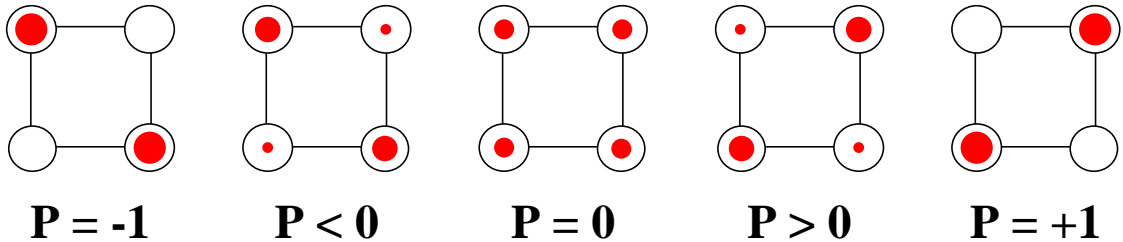

P = -1        P < 0        P = 0        P > 0        P = +1

Figure 1. Cell polarization is treated as a continuous variable.

localize on a particular dot but can tunnel between dots. No tunneling occurs between cells. The polarization P of the cell is defined from the expectation values of the charge on each dot.

$$P \equiv \frac{(\rho_1 + \rho_3) - (\rho_2 + \rho_4)}{\rho_1 + \rho_2 + \rho_3 + \rho_4} \quad (3)$$

P can vary continuously between -1 and +1 as shown in the figure. We describe the quantum state of a cell using two basis states $|\phi_1\rangle$ and $|\phi_2\rangle$ which are completely polarized.

$$|\psi\rangle = \alpha|\phi_1\rangle + \beta|\phi_2\rangle \qquad \psi = \begin{bmatrix} \alpha \\ \beta \end{bmatrix} \quad (4)$$

Using these two components the cell polarization is given by

$$P = |\alpha|^2 - |\beta|^2 . \quad (5)$$

The Coulomb interaction between adjacent cells increases the energy of the configuration if the two cell polarizations differ. This can be account for by including an energy shift corresponding to the weighted sum of the neighboring polarizations. We define this weighted sum P as follows;

$$P_\kappa = \sum_{\lambda \in S_\kappa} w(\mathbf{R}_\kappa - \mathbf{R}_\lambda) P_\lambda . \quad (6)$$

where the sum is over an appropriate neighborhood $S_\kappa$ about cell $\kappa$. The Hamiltonian for each cell can then be written as

$$H_\kappa = \begin{bmatrix} E_0 - \frac{1}{2}\overline{P}_\kappa E_k & -\gamma \\ -\gamma & E_0 + \frac{1}{2}\overline{P}_\kappa E_k \end{bmatrix} \quad (7)$$

where $\gamma$ is the interdot tunneling energy and $E_k$ is the electrostatic energy cost of two adjacent fully polarized cells having opposite polarization as shown in Figure 2. If we assume that there are no quantum entanglements between cells, then the dynamics of the array is simply given by a set of coupled Schrödinger equations for each cell.

$$i\hbar \frac{\partial}{\partial t}|\psi_\kappa\rangle = H_\kappa |\psi_\kappa\rangle \quad (8)$$

This approximation treats exchange and correlation effects exactly within each cell (for the model) and treats intercellular interactions at the level of Hartree-Fock. Allowing correlation effects that produced mixed intercellular states would makes connecting to a CNN description impossible because of the need for local cell state information. Moreover, in our simulations of dynamic switching of cellular arrays we found that including the correlations between cells did not alter the qualitative behavior (though it did increase the speed of the intercellular responses.)

## IV. Formulating quantum dynamics as CNN dynamics

To transform the quantum mechanical description of an array into a CNN-style description the first step is to reduce the number of local dynamical variables describing each cell. The two-state approximation of equation (4) requires two complex numbers, $\alpha$ and $\beta$, to describe a state. This entails four real degrees of freedom. One degree of freedom can be removed by noting that the overall quantum phase of the state is arbitrary (again here the condition of no intercellular mixed states is required). A second degree of freedom is removed by using the normalization condition

$$1 = |\alpha|^2 + |\beta|^2 \quad . \tag{9}$$

It is then possible to rewrite the state description in terms of two real degrees of freedom, P and $\phi$.

$$\psi = \begin{bmatrix} \sqrt{\dfrac{1+P}{2}} \\ \sqrt{\dfrac{1-P}{2}} e^{i\varphi} \end{bmatrix} \tag{10}$$

Notice that P represents a classical degree of freedom — it is related to expectation values of observables. By contrast $\phi$ is a fundamentally quantum variable, a quantum mechanical phase. The dynamical equations derived from the Schrödinger equation (8) can be rewritten as equations for P and $\phi$.

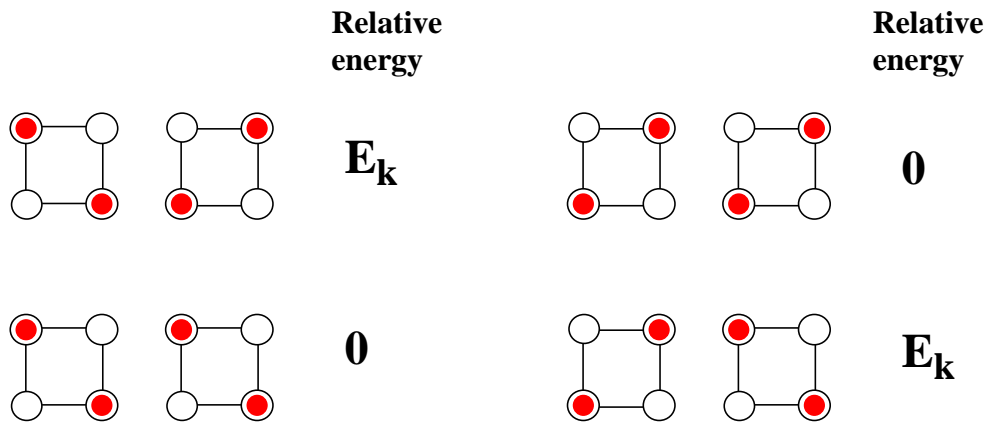

Figure 2. The energy cost of neighboring cells having opposite polarization.

$$\hbar \frac{\partial}{\partial t} P = -2\gamma \sqrt{1 - P^2} \sin \varphi \tag{11}$$

$$\hbar \frac{\partial}{\partial t} \varphi = -\overline{P} E_k + 2\gamma \frac{P}{\sqrt{1 - P^2}} \cos \varphi \tag{12}$$

Equations (11) and (12) are the Q-CNN state equation, analogous to equation (2). We can see by comparing (12) with the Hamiltonian (7) that the synaptic law is given by:

$$I_\lambda^s = E_k \overline{P}_\lambda = \sum_{\lambda \in S_\kappa} w(\mathbf{R}_\kappa - \mathbf{R}_\lambda) P_\lambda \tag{13}$$

We have shown [1] that a line of cells has a stable self-polarization at a value $P = \pm P_{sat}$ where $P_{sat}$ is determined by the intercellular Coulomb coupling and the tunneling. Using (11) and (12) we can find a closed form expression for $P_{sat}$.

$$P_{sat} = \sqrt{1 - \left(\frac{\gamma}{E_k}\right)^2} \tag{14}$$

We have previously calculated the properties of a line of cells using a complete many-particle basis consisting of 25 states per cell We examined the line both with and without intercellular correlations [5]. The primary feature of interest was the propagation of a switched pulse along the line. *A priori* it is not obvious that a treatment as simple as the two state model we describe here is sufficient to capture this behavior. The solutions to the dynamic equation shown in Figure 3 demonstrates that it does indeed. A pulse is seen to propagate down the line. If we neglect the quantum mechanical dynamical variable φ, this propagation does not occur. It can be seen from the figure as well as from equation (11) that the sign of φ determines the time derivative of P and thus the direction of wave propagation.

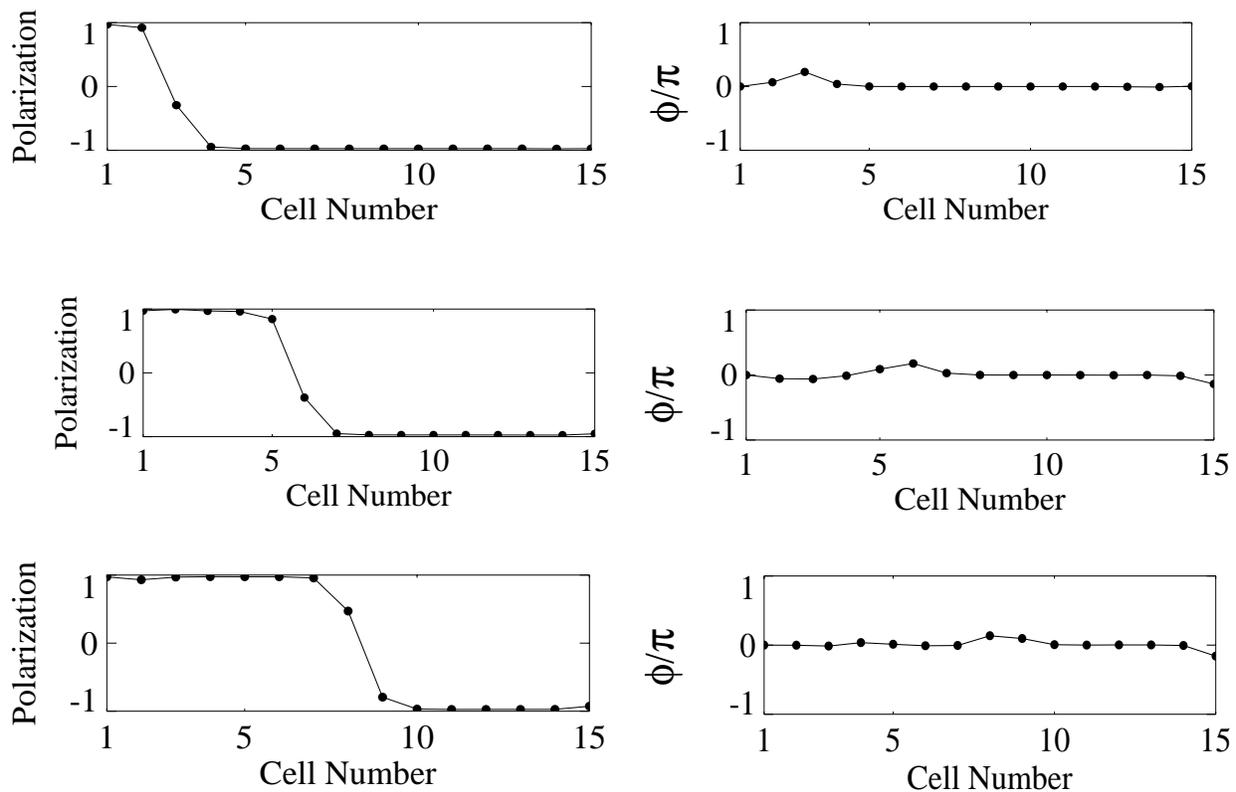

Figure 3. Wave front motion in a simple Q-CNN. The first cell is switched abruptly from -$P_{sat}$ to $P_{sat}$. Snapshots of P and $\phi$ for the line show the line of cells switching as the pulse moves from left to right.

## V. Generalization of Quantum Cellular Neural Networks

Although we have employed a fairly simple model for demonstrating Q-CNN behavior, the general features of the paradigm are clear.

1. Each cell is a quantum system. The specification of the quantum system can distinguish $N_c$ classical degrees of freedom and $N_q$ quantum degrees of freedom.

2. The interaction between cells, the synaptic input, depends only on the classical degrees of freedom. This corresponds to an intercellular Hartree-Fock approximation. The precise form of the synaptic law is determined by the physics of the intercellular interaction.

3. The state equations are derived from the time-dependent Schrödinger equation. One state equation exists for each classical and quantum degree of freedom.

It is notable that the classical degrees of freedom carry the information from cell to cell but the quantum degrees of freedom are necessary to carry information from the one time to the next.

This can be seen in the example shown in Figure 3 for which the direction of pulse propagation is encoded in the phase variable.

## VI. Conclusions

We have defined the Q-CNN paradigm and examined it in the case of a simple two-state model of the cell. The system is sufficiently rich to reproduce the wave propagation behavior seen in a fuller quantum treatment. The general features of Q-CNN architecture have been outlined. Of particular interest is the distinction between information-bearing classical degrees of freedom and quantum degrees of freedom which are necessary for proper temporal evolution.

In making the connection between coupled quantum cells and the existing CNN paradigm we have made the first, very preliminary step in appropriating the results of work in classical CNN circuit theory for use in quantum device applications. Further investigation of 2D is the next essential step.

### Acknowledgments

This work was supported by the Office of Naval Research and the Defense Advanced Research Projects Agency.